\pdfoutput=1
\documentclass[a4paper, 11pt]{article}
\usepackage{newtxtext}
\usepackage[libertine]{newtxmath}
\usepackage{microtype}
\usepackage{cite}
\usepackage{color}
\definecolor{color1}{rgb}{0, 0, 0.5}
\usepackage[colorlinks,allcolors=color1]{hyperref}
\usepackage[top=20truemm, bottom=20truemm,
left=25truemm, right=25truemm]{geometry}

\renewcommand{\d}{\text{d}}
\renewcommand{\i}{\text{i}}
\DeclareMathOperator{\Tr}{Tr}
\allowdisplaybreaks[1]

\title{Canonical analysis of covariant unimodular gravity\\
  and an extension of the Kodama state
}
\author{
  Shinji Yamashita\footnote{
    \texttt{shinji0yamashita@gmail.com}
  }
  \\
  {\small \textit{
      National Institute of Technology, Niihama College,
      Ehime, Japan}
  }\\
}
\date{}

\begin{document}
\maketitle
\begin{abstract}
  We carry out the canonical analysis of a covariant version of
  unimodular gravity in terms of the connection representation.
  We then proceed to quantize this theory by implementing the Dirac
  procedure.
  We confirm whether and how the Kodama state, which is a solution of
  quantum general relativity, can be extended into covariant
  unimodular gravity.
  Finally, we discuss the difference of quantum states between
  covariant unimodular gravity, the original unimodular
  gravity, and general relativity.
\end{abstract}

\section{Introduction}

\label{sec:introduction}
Unimodular gravity is a simple modification of general relativity
(GR) in which the determinant of the spacetime metric is restricted
to be constant.
Due to this restriction, unimodular gravity does not preserve the
full diffeomorphism invariance.
Nevertheless,
the classical field equations in unimodular
gravity are almost the same as in GR.
A subtle but crucial difference from GR is that the cosmological
constant is treated as an arbitrary integration constant
\cite{Bij1982}.
This arbitrariness brings a different perspective on the cosmological
constant problem \cite{Unruh1989,Smolin2009}.

The Hamiltonian analysis of unimodular gravity in terms of the
Arnowitt-Deser-Misner (ADM) variables has been performed in
Refs.~\cite{Kluson2015,Bufalo2015}.
In contrast to ordinary GR, the lapse function is not regarded as an
independent variable due to the unimodular condition, and the
Hamiltonian constraint is a second-class constraint.
Additionally, the total Hamiltonian does not vanish on the constraint
surface.
In canonical quantum theory, these differences from GR cause the
differences in the physical states.
Specifically, the physical state of unimodular gravity is constructed
from the eigenstates of the cosmological constant.
Furthermore, unimodular gravity can have an appropriate time variable,
and the physical state obeys the Schr\"odinger-like equation rather
than the Wheeler-DeWitt one.
In this sense, one can avoid the problem of time in quantum gravity
\cite{Unruh1989,Unruh1989a}.

In this paper, we perform the canonical analysis of unimodular gravity
and its quantization; however, the theory we will discuss has two
different points from the original unimodular gravity explained above.
The first point is that we employ a covariant version of unimodular
gravity that was suggested in Ref.~\cite{Henneaux1989}.
In this framework, the square root of the determinant of the spacetime
metric is equal to the divergence of a densitized vector field, and
the full diffeomorphism invariance is retained.
Moreover, one can introduce time as spacetime volume \cite{Brown1989}.
This theory gives the same physics as the original unimodular gravity
at least at the classical level, while we can expect that these two
unimodular theories provide different quantum theories because of the
difference of the constraints.

The second point is that we describe the theory in terms of the
connection representation instead of the ADM one.
Within the framework, one of the configuration variables is the
Ashtekar-Barbero connection with the Barbero-Immirzi parameter
$\beta$, and its conjugate momentum is the densitized triad
\cite{Ashtekar1987,Barbero1995,Ashtekar2004}.
The advantage of this representation is that the constraints are
somewhat simpler than those in the ADM representation.
Thanks to the simplicity, several solutions that satisfy quantum
first-class constraints of GR have been found.
In particular, the Kodama state, which is also called the
Chern-Simons state, is a well-known solution of quantum GR with a
nonvanishing cosmological constant in the case of $\beta=\i$ (the
imaginary unit) \cite{Kodama1990,Smolin2002}.
A generalization of the Kodama state for real values of $\beta$ was
also suggested in Ref.~\cite{Randono2006}.
On the other hand, the Kodama state is not regarded as a physical
state in the original unimodular gravity \cite{Yamashita2020}.

The aim of this paper is to confirm the difference between
covariant unimodular gravity, the original unimodular gravity, and GR,
especially at the quantum level.
The manuscript is organized as follows.
In Sec.~\ref{sec:canonical_analysis}, we perform the Hamiltonian
analysis of covariant unimodular gravity in terms of the connection
representation.
While some work along this line has been done for $\beta=\i$
\cite{Bombelli1991,Smolin2011}, we further develop the theory for real
values of $\beta$.
Although this choice makes the Hamiltonian constraint more complicated
than in the case of $\beta=\i$, it facilitates the construction of the
inner product in quantum theory.
In Sec.~\ref{sec:quantization}, we quantize the theory by applying
the Dirac procedure \cite{Henneaux1992,Matschull1996}.
Then, we consider whether and how the Kodama state can be extended
into covariant unimodular gravity, mainly following
Ref.~\cite{Randono2006}.
In Sec.~\ref{sec:conclusions}, we summarize our results including the
comparison between covariant unimodular gravity, the original
unimodular gravity, and GR.

We use the following notation.
Greek letters $\mu, \nu, \dots \in \{\tau, 1, 2, 3\}$ indicate
four-dimensional spacetime indices where $\tau$ is the time flow
component.
Capital letters $I, J, \dots \in \{0, 1, 2, 3\}$ are Lorentz indices.
Letters $a, b, \dots \in \{1, 2, 3\}$ are three-dimensional spatial
indices, and $i, j, \dots \in \{1, 2, 3\}$ are internal su(2) Lie
algebra indices.
We employ a four-metric signature $(-, +, +, +)$, and use units in
which the speed of light is unity.

\section{Canonical analysis}
\label{sec:canonical_analysis}

The simplest action of the original unimodular gravity without matter
is obtained by modifying the Einstein-Hilbert action
\begin{align}
  S_{\text{UG}}(g_{\mu\nu}, \Lambda)
  & = \frac{1}{2k}\int \d^{4}x\
    \left[ \sqrt{-\det g}\ R^{(4)}
    - \Lambda \left( \sqrt{-\det g} - \alpha \right)\right],
    \label{eq:1}
\end{align}
where $k$ is Newton's constant times $8\pi$, $R^{(4)}$ is a scalar
curvature of four-dimensional spacetime, $\Lambda$ is a scalar field
that plays the role of a Lagrange multiplier, and $\alpha$ is a fixed
scalar density.
The variation with respect to $\Lambda$ leads to the unimodular
condition $\sqrt{-\det g} - \alpha = 0$.
In this framework, the spacetime diffeomorphism is restricted so that
the value of the determinant of the four-metric remains unchanged.

The reformulation of unimodular gravity that ensures full
diffeomorphism invariance was introduced by Henneaux and Teitelboim
\cite{Henneaux1989}.
The action has the form
\begin{align}
  S_{\text{HT}}(g_{\mu\nu}, \Lambda, \phi^{\mu})
  = \frac{1}{2k}\int \d^{4}x\
  \left[ \sqrt{-\det g}\ R^{(4)}
  - \Lambda \left( \sqrt{-\det g} - \partial_{\mu}\phi^{\mu} \right)
  \right],
  \label{eq:2}
\end{align}
where $\phi^{\mu}$ is a densitized vector field of weight one.
The unimodular condition is rewritten as
$\sqrt{-\det g}\ - \partial_{\mu}\phi^{\mu} = 0$.
Let us write the action corresponding to \eqref{eq:2} in terms of the
connection representation.
This can be done by modifying the Holst action \cite{Holst1996}
\begin{align}
  &S(\omega_{\mu}^{IJ}, e_{I}^{\mu}, \Lambda, \phi^{\mu})
    =\int \d^{4}x\ L
    \notag \\
  & = -\frac{1}{2k\beta}\int
    e^{I} \wedge e^{J} \wedge
    \left( R_{IJ}^{(4)} -\frac{\beta}{2}\epsilon_{IJKL} R^{(4)KL}
    \right)
    - \frac{1}{48k} \int \Lambda \epsilon_{IJKL}
    e^{I} \wedge e^{J} \wedge e^{K} \wedge e^{L}
    + \frac{1}{2k} \int \d^{4}x\ \Lambda \partial_{\mu}\phi^{\mu},
    \label{eq:3}
\end{align}
where $e^{I}$ is a cotetrad,
$R^{(4)IJ} = \d \omega^{IJ} + \omega^{I}{}_{K} \wedge \omega^{KJ}$
is a curvature of the spin connection $\omega_{\mu}^{IJ}$,
and $\beta$ is the Barbero-Immirzi parameter that takes nonvanishing
real values.
The $3 + 1$ form of the above action under the time gauge
$e_{a}^{0}=0$ is written as
\begin{align}
  S= \frac{1}{k\beta}\int \d^{4}x\
  \left(
  E_{i}^{a}\dot{A}_{a}^{i}
  - A_{\tau}^{i}G_{i} - N^{a}V_{a} - NC
  \right)
  + \frac{1}{2k}\int \d^{4}x
  \left(\Lambda \dot{\phi}^{\tau}
  - \Lambda N \det e
  - \phi^{a}\partial_{a}\Lambda \right),
  \label{eq:4}
\end{align}
where $\det e$ is a determinant of a cotriad $e_{a}^{i}$,
$E_{i}^{a} = (\det e) e_{i}^{a}$ is a densitized triad,
$A_{\mu}^{i} = -\frac{1}{2}\epsilon^{i}{}_{jk}\omega_{\mu}^{jk}
- \beta\omega_{\mu}^{0i}$, $N^{a}$  is a shift vector, and $N$ is a
lapse function.
The spatial component of $A_{\mu}^{i}$ is usually expressed as
$A_{a}^{i} = \Gamma_{a}^{i} + \beta K_{a}^{i}$, where
$\Gamma_{a}^{i} = -\frac{1}{2}\epsilon^{i}{}_{jk}\omega_{a}^{jk}$
is a three-dimensional spin connection compatible with $e_{i}^{a}$,
and $K_{a}^{i} = -\omega_{a}^{0i} = K_{ab}e_{j}^{b}\delta^{ij}$
is related with the extrinsic curvature $K_{ab}$ and $e_{i}^{a}$.
In addition,
\begin{align}
  G_{i}
  &= - \left( D_{a}E^{a} \right)_{i}
    = -\left( \partial_{a}E_{i}^{a}
    + \epsilon_{ij}{}^{k}A_{a}^{j}E_{k}^{a} \right),
    \label{eq:5}
  \\
  V_{a}
  &= - E_{i}^{b}F_{ba}^{i},
    \label{eq:6}
  \\
  C
  &= \frac{1}{2 \beta \sqrt{\det E}}\epsilon^{ijk}
    E_{i}^{a}E_{j}^{b}
    \left[ \left( 1+\beta^{2} \right) R_{abk}(E) - F_{abk} \right],
    \label{eq:7}
\end{align}
where $F_{ab}^{i} = \partial_{a}A_{b}^{i} - \partial_{b}A_{a}^{i}
+ \epsilon^{i}{}_{jk}A_{a}^{j}A_{b}^{k}$
is a curvature of $A_{a}^{i}$,
$R_{ab}^{i}(E) = \partial_{a}\Gamma_{b}^{i} -
\partial_{b}\Gamma_{a}^{i} +
\epsilon^{i}{}_{jk}\Gamma_{a}^{j}\Gamma_{b}^{k}$ is a curvature
of $\Gamma_{a}^{i}$ that is constructed from $E_{i}^{a}$,
and $\det E = (\det e)^{2}$ is a determinant of $E_{i}^{a}$.
Note that while $C$ \eqref{eq:7} is often expressed as
\begin{align}
  C = \frac{\beta}{2\sqrt{\det E}}\epsilon^{ijk}E_{i}^{a}E_{j}^{b}
  \left[
  F_{abk} - \left( 1 + \beta^{2} \right)
  \epsilon_{klm}K_{a}^{l}K_{b}^{m}
  \right],
  \label{eq:8}
\end{align}
we use the former expression \eqref{eq:7} for latter convenience.

The configuration variables of this theory are
$\left( A_{\tau}^{i}, A_{a}^{i}, N, N^{a}, \Lambda,
  \phi^{\tau}, \phi^{a} \right)$.
The canonical conjugate momenta (multiplied by $k\beta$ or $2k$) are
given by
\begin{align}
  \pi_{i}
  &= k\beta \frac{\partial L}{\partial \dot{A}_{\tau}^{i}} = 0,
  &
    E_{i}^{a}
  &= k\beta \frac{\partial L}{\partial \dot{A}_{a}^{i}},
    \label{eq:9}
  \\
  \pi_{N}
  &=  k\beta \frac{\partial L}{\partial \dot{N}} = 0,
  &
    \pi_{a}
  &=  k\beta \frac{\partial L}{\partial \dot{N}^{a}} = 0,
    \label{eq:10}
  \\
  \pi_{\Lambda}
  &=2k \frac{\partial L}{\partial \dot{\Lambda}} = 0,
  &&
     \label{eq:11}\\
  p_{\tau}
  &=2k \frac{\partial L}{\partial \dot{\phi^{\tau}}} = \Lambda,
  &
    p_{a}
  &=2k \frac{\partial L}{\partial \dot{\phi^{a}}} = 0.
    \label{eq:12}
\end{align}
These momenta yield primary constraints
\begin{align}
  \pi_{i}\approx \pi_{N} \approx \pi_{a} \approx \pi_{\Lambda}
  \approx \Pi\approx p_{a} \approx 0,
  \label{eq:13}
\end{align}
where $\Pi=p_{\tau}-\Lambda$, and the symbol $\approx$ is weak
equality, which indicates that the equality holds on the constraint
surface.
The fundamental Poisson bracket relations are
\begin{align}
  \begin{aligned}
    \left\{ A_{\tau}^{i}(x), \pi_{j}(y) \right\} &=k\beta
    \delta_{j}^{i}\delta^{3}(x-y), & \left\{ A_{a}^{i}(x),
      E_{j}^{b}(y) \right\} &=k\beta
    \delta_{a}^{b}\delta_{j}^{i}\delta^{3}(x-y),
    \\
    \left\{ N(x), \pi_{N}(y) \right\} &=k\beta \delta^{3}(x-y), &
    \left\{ N^{a}(x), \pi_{b}(y) \right\} &=k\beta
    \delta_{b}^{a}\delta^{3}(x-y),
    \\
    \left\{ \Lambda(x), \pi_{\Lambda}(y) \right\} &=2k
    \delta^{3}(x-y),
    \\
    \left\{ \phi^{\tau}(x), p_{\tau}(y) \right\} &= 2k\delta^{3}(x-y),
    & \left\{ \phi^{a}(x), p_{b}(y) \right\} &=
    2k\delta_{b}^{a}\delta^{3}(x-y).
  \end{aligned}
                                               \label{eq:14}
\end{align}
The total Hamiltonian $H_{\text{T}}$ is a combination of the ordinary
Hamiltonian and the primary constraints with Lagrange multipliers
$v^{i}, v^{a}, v_{N}, v_{\Lambda}, \tilde{w}, \tilde{w}^{a}$:
\begin{align}
  &H_{\text{T}}(A_{\tau}^{i},\pi_{i},A_{a}^{i},E_{i}^{a},
    N, \pi_{N}, N^{a}, \pi_{a}, \Lambda, \pi_{\Lambda},
    \phi^{\tau}, p_{\tau}, \phi^{a},p_{a})
    \notag \\
  &= \int \d^{3}x \biggl[
    \frac{1}{k\beta}\left( A_{\tau}^{i}G_{i} + N^{a}V_{a} + NC \right)
    + \frac{1}{2k}
    \left( \Lambda N \det e + \phi^{a}\partial_{a}\Lambda \right)
    \notag \\
  &\hspace{5em}
    + \frac{1}{k\beta}
    \left( v^{i}\pi_{i} + v^{a}\pi_{a} + v_{N}\pi_{N} \right)
    + \frac{1}{2k}
    \left( v_{\Lambda}\pi_{\Lambda} + \tilde{w}\Pi
    + \tilde{w}^{a}p_{a} \right)
    \biggr],
    \label{eq:15}
\end{align}
where $\tilde{w}$ and $\tilde{w}^{a}$ are densities of weight one.
In general, every constraint must satisfy the stability condition,
that is, each constraint must hold under time evolution on the
constraint surface.
Applying this condition to the primary constraints \eqref{eq:13}, we
have
\begin{align}
  \left\{\pi_{i}, H_{\text{T}}\right\}
  &= - G_{i} \approx 0,
    \label{eq:16}
  \\
  \left\{\pi_{a}, H_{\text{T}}\right\}
  &= - V_{a} \approx 0,
    \label{eq:17}
  \\
  \left\{\pi_{N}, H_{\text{T}}\right\}
  &= - \Phi \approx 0,
    \label{eq:18}
  \\
  \left\{\pi_{\Lambda}, H_{\text{T}}\right\}
  &= -\left(N\det e - \partial_{a}\phi^{a} - \tilde{w}\right)
    \approx 0,
    \label{eq:19}
  \\
  \left\{\Pi, H_{\text{T}}\right\}
  & = -v_{\Lambda} \approx 0,
    \label{eq:20}
  \\
  \left\{p_{a}, H_{\text{T}}\right\}
  & = -\Sigma_{a} \approx 0,
    \label{eq:21}
\end{align}
where
\begin{align}
  \Phi
  &= \frac{1}{2\beta \sqrt{\det E}} \epsilon^{ijk}E_{i}^{a}E_{j}^{b}
    \left[
    \left( 1 + \beta^{2} \right) R_{abk} - F_{abk}
    + \frac{\beta^{2}\Lambda}{6}\epsilon_{abc}E_{k}^{c}
    \right],
    \label{eq:22}
  \\
  \Sigma_{a}
  &= \partial_{a}\Lambda.
    \label{eq:23}
\end{align}
While $v^{i}, v^{a}, v_{N}$, and $\tilde{w}^{a}$ remain unspecified,
$v_{\Lambda}$ and $\tilde{w}$ are determined by Eqs.~\eqref{eq:19} and
\eqref{eq:20} as
\begin{align}
  v_{\Lambda}& = 0,
  & \tilde{w}
  &= N\det e - \partial_{a}\phi^{a}.
    \label{eq:24}
\end{align}
The secondary constraints $G_{i}\approx 0$~\eqref{eq:16},
$V_{a}\approx 0$~\eqref{eq:17}, and $\Phi \approx 0$~\eqref{eq:18}
are the Gauss, vector, and Hamiltonian constraints, respectively,
which are the same as those in GR.
The secondary constraint $\Sigma_{a} \approx 0$~\eqref{eq:21}
implies that $\Lambda$ is a spatial constant.
We define the smeared versions of the secondary constraints:
\begin{align}
  G[X^{i}]
  &= \frac{1}{k\beta}\int \d^{3}x\ X^{i}G_{i}(x),
    \label{eq:25}
  \\
  V[X^{a}]
  &= \frac{1}{k\beta}\int \d^{3}x\ X^{a}V_{a}(x),
    \label{eq:26}
  \\
  \Phi[X]
  &= \frac{1}{k\beta}\int \d^{3}x\ X\Phi(x),
    \label{eq:27}
  \\
  \Sigma[\tilde{X}^{a}]
  &= \frac{1}{2k}\int \d^{3}x\ \tilde{X}^{a}\Sigma_{a}(x),
    \label{eq:28}
\end{align}
where $X^{i}, X^{a}, X$ are test functions, and $\tilde{X}^{a}$ is a
densitized test function.
One can check that every secondary constraint has a weakly vanishing
Poisson bracket with the total Hamiltonian \eqref{eq:15}.
Then, every secondary constraint automatically satisfies the stability
condition, and no more constraints arise.

Now, we can classify the primary constraints
$\left(
  \pi_{i}, \pi_{a}, \pi_{N}, \pi_{\Lambda}, \Pi, p_{a}
\right)$
and the secondary constraints
$\left( G[X^{i}], V[X^{a}], \Phi[X], \Sigma[\tilde{X}^{a}] \right)$
into the first- and second-class constraints.
The weakly nonvanishing Poisson brackets among these constraints are
\begin{align}
  \left\{ \pi_{\Lambda}(x), \Pi(y) \right\}
  &= 2k \delta^{3}(x-y),
    \label{eq:29}
  \\
  \left\{ \pi_{\Lambda}, \Phi[X] \right\}
  &= -X\det e,
    \label{eq:30}
  \\
  \left\{ \pi_{\Lambda}, \Sigma[\tilde{X}^{a}] \right\}
  &= \partial_{a}\tilde{X}^{a}.
    \label{eq:31}
\end{align}
Hence constraints
$\left(  \pi_{\Lambda}, \Pi, \Phi[X], \Sigma[\tilde{X}^{a}] \right)$
are second class, and the remaining constraints are first class.

The number of the second-class constraints can be reduced by replacing
$\Phi[X]$ and $\Sigma[\tilde{X}^{a}]$ with the modified constraints
$\Phi'[X]$ and $\Sigma'[\tilde{X}^{a}]$, respectively:
\begin{align}
  \Phi'[X]
  &= \Phi[X] + \frac{1}{2k}\int \d^{3}x\ X \Pi \det e
    \notag \\
  &= \frac{1}{2k\beta^{2}}\int \d^{3}x\
    \frac{X}{\sqrt{\det E}}\epsilon^{ijk}E_{i}^{a}E_{j}^{b}
    \left[
    \left( 1 + \beta^{2} \right) R_{abk} - F_{abk}
    + \frac{\beta^{2}}{6}\epsilon_{abc}p_{\tau}E_{k}^{c}
    \right],
    \label{eq:32}
  \\
  \Sigma'[\tilde{X}^{a}]
  &= \Sigma[\tilde{X}^{a}]
    + \frac{1}{2k}\int \d^{3}x\ \tilde{X}^{a} \partial_{a}\Pi
    =\frac{1}{2k}\int \d^{3}x\ \tilde{X}^{a}\partial_{a}p_{\tau}.
    \label{eq:33}
\end{align}
In fact, the weakly nonvanishing Poisson bracket among the primary
constraints
$\left( \pi_{i}, \pi_{a}, \pi_{N}, \pi_{\Lambda}, \Pi, p_{a} \right)$
and the secondary constraints
$\left( G[X^{i}], V[X^{a}], \Phi'[X], \Sigma'[\tilde{X}^{a}] \right)$
is only one:
\begin{align}
  \left\{ \pi_{\Lambda}(x), \Pi(y) \right\}
  =2k \delta^{3}(x-y).
  \label{eq:34}
\end{align}
Then, constraints $\left( \pi_{\Lambda}, \Pi \right)$ are second
class, and the remaining constraints are first class.
Let us count the degrees of freedom of this theory.
Variables
$\left( A_{\tau}^{i}, A_{a}^{i}, N, N^{a}, \Lambda, \phi^{\tau},
  \phi^{a}\right)$ have $3+9+1+3+1+1+3=21$ components.
The first-class constraints
$\left( \pi_{i}, \pi_{a}, \pi_{N}, p_{a}, G[X^{i}], V[X^{a}],
  \Phi'[X], \Sigma'[\tilde{X}^{a}] \right)$ constrain
$3+3+1+3+3+3+1+1=18$ components.
Note that $\Sigma'[\tilde{X}^{a}]$ constrains only one component,
because this constraint is parametrized by $\partial_{a}\tilde{X}^{a}$
rather than $\tilde{X}^{a}$.
The second-class constraints $\left( \pi_{\Lambda}, \Pi \right)$
constrain $(1+1)/2=1$ component.
Therefore, the local degrees of freedom in configuration space are
$21-18-1=2$.
This result is consistent with GR and the original unimodular gravity
\cite{Yamashita2020}.

Using the second-class constraints $\Pi\approx 0$ and
$\pi_{\Lambda} \approx 0$, we can eliminate variables $\Lambda$ and
$\pi_{\Lambda}$ by substituting
\begin{align}
  \Lambda
  &= p_{\tau},
  &
    \pi_{\Lambda}
  &= 0.
    \label{eq:35}
\end{align}
After the elimination, the total Hamiltonian \eqref{eq:15} is
rewritten as
\begin{align}
  &H_{\text{T}}
    (A_{\tau}^{i},\pi_{i},A_{a}^{i},E_{i}^{a},
    N, \pi_{N}, N^{a}, \pi_{a}, \phi^{\tau}, p_{\tau}, \phi^{a},p_{a})
    \notag \\
  &= G[A_{\tau}^{i}] + V[N^{a}] + \Phi'[N] + \Sigma'[\phi^{a}]
    + \int \d^{3}x\
    \left[
    \frac{1}{k\beta}
    \left(v^{i}\pi_{i} + v^{a}\pi_{a} + v_{N}\pi_{N}\right)
    + \frac{1}{2k} \tilde{w}^{a}p_{a}
    \right].
    \label{eq:36}
\end{align}
The constraint $\Sigma'[\tilde{X}^{a}] \approx 0$ \eqref{eq:33}
and the evolution equation
$\left\{ p_{\tau}, H_{\text{T}} \right\} \approx0$
imply that $p_{\tau}$ is a spacetime constant.
In addition, the evolution equation
$\left\{ \phi^{\tau}, H_{\text{T}} \right\}
=N\det e - \partial_{a}\phi^{a}$
leads to the covariant version of the unimodular condition
$N\det e - \partial_{\mu}\phi^{\mu}=0$.
Obviously, $p_{\tau}$ and $\Lambda$ correspond to the cosmological
constant (times two) in GR.

We can introduce the spatial diffeomorphism constraint
$\mathcal{D}[X^{a}]$ by extending the vector constraint $V[X^{a}]$:
\begin{align}
  \mathcal{D}[X^{a}]
  = V[X^{a}] + G[X^{a}A_{a}^{i}] - \Sigma'[X^{a}\phi^{\tau}].
  \label{eq:37}
\end{align}
This constraint generates the spatial diffeomorphism for the dynamical
variables
\begin{align}
  \left\{ A_{a}^{i}, \mathcal{D}[X^{b}] \right\}
  &= \mathcal{L}_{\vec{X}}A_{a}^{i},
  &
    \left\{ E_{i}^{a}, \mathcal{D}[X^{b}] \right\}
  &= \mathcal{L}_{\vec{X}} E_{i}^{a},
    \label{eq:38}
  \\
  \left\{ \phi^{\tau}, \mathcal{D}[X^{a}] \right\}
  &= \mathcal{L}_{\vec{X}}\phi^{\tau},
  &
    \left\{ p_{\tau}, \mathcal{D}[X^{a}] \right\}
  &= \mathcal{L}_{\vec{X}} p_{\tau},
    \label{eq:39}
\end{align}
where $\mathcal{L}_{\vec{X}}$ is a Lie derivative along $\vec{X}$.
The constraint $\mathcal{D}[X^{a}]$ is first class, and holds the
stability condition
\begin{align}
  \left\{ \mathcal{D}[X^{a}], H_{\text{T}} \right\}
  = G\left[ \mathcal{L}_{\vec{X}}A_{\tau}^{i} \right]
  + V\left[ \mathcal{L}_{\vec{X}}N^{a} \right]
  + \Phi'\left[ \mathcal{L}_{\vec{X}}N \right]
  + \Sigma'\left[ \mathcal{L}_{\vec{X}}\phi^{a} \right]
  \approx 0.
  \label{eq:40}
\end{align}
We employ $\mathcal{D}[X^{a}]$ instead of $V[X^{a}]$ as an element of
the first-class constraints.

\section{Quantization}
\label{sec:quantization}

After the reduction \eqref{eq:35}, all the remaining constraints
$\left( \pi_{i}, \pi_{a}, \pi_{N}, p_{a}, G[X^{i}],
  \mathcal{D}[X^{a}], \Phi'[X], \Sigma'[\tilde{X}^{a}] \right)$
belong to first class.
Hence, we can proceed to quantize this theory by replacing Poisson
brackets $\left\{ \bullet, \bullet \right\}$ with quantum commutators
$(\i \hbar)^{-1}\left[ \hat{\bullet}, \hat{\bullet} \right]$.
The quantum operators corresponding to the canonical variables are
given by
\begin{align}
  \hat{A}_{\tau}^{i} &= A_{\tau}^{i},
  &
    \hat{\pi}_{i}
  &=-\i\hbar k\beta\frac{\delta}{\delta A_{\tau}^{i}},
    \label{eq:41}
  \\
  \hat{A}_{a}^{i} &= A_{a}^{i},
  &
    \hat{E}_{i}^{a}
  &= -\i\hbar k\beta \frac{\delta}{\delta A_{a}^{i}},
    \label{eq:42}
  \\
  \hat{N} &= N,
  &
    \hat{\pi}_{N}
  &= -\i \hbar k\beta \frac{\delta}{\delta N},
    \label{eq:43}
  \\
  \hat{N}^{a} &= N^{a},
  &
    \hat{\pi}_{a}
  &= -\i \hbar k\beta \frac{\delta}{\delta N^{a}},
    \label{eq:44}
  \\
  \hat{\phi}^{\tau} &= \phi^{\tau},
  &
    \hat{p}_{\tau}
  &= -2\i\hbar k \frac{\delta}{\delta \phi^{\tau}},
    \label{eq:45}
  \\
  \hat{\phi}^{a} &= \phi^{a},
  &
    \hat{p}_{a}
  &= -2\i\hbar k \frac{\delta}{\delta \phi^{a}}.
    \label{eq:46}
\end{align}
A physical state $\Psi$ must satisfy the quantized first-class
constraints
\begin{align}
  \hat{\pi}_{i} \Psi
  &= \hat{\pi}_{N} \Psi=\hat{\pi}_{a} \Psi = \hat{p}_{a} \Psi = 0,
    \label{eq:47}
  \\
  \hat{G}[X^{i}] \Psi
  &= \hat{\mathcal{D}}[X^{a}] \Psi
    = \hat{\Phi}'[X] \Psi
    = \hat{\Sigma}'[\tilde{X}^{a}] \Psi = 0.
    \label{eq:48}
\end{align}
Constraints \eqref{eq:47} indicate that $\Psi$ should be independent
from $A_{\tau}^{i}, N, N^{a}$, and $\phi^{a}$, that is,
\begin{align}
  \Psi = \Psi[\phi^{\tau}, A_{a}^{i}].
  \label{eq:49}
\end{align}
We assume that the wave functional is variable-separable, namely,
$\Psi[\phi^{\tau}, A_{a}^{i}] = \Psi[\phi^{\tau}] \Psi[A_{a}^{i}]$.
Constraint $\hat{\Sigma}'[\tilde{X}^{a}]\Psi =
(2k)^{-1}\int \d^{3}x\ \tilde{X}^{a} \partial_{a}\hat{p}_{\tau}\Psi
\approx 0$ implies that $\Psi$ has the form
\begin{align}
  \Psi[\phi^{\tau}, A_{a}^{i}]
  = \exp\left[
  \frac{\lambda}{-2\i\hbar k} \int \d^{3}x\ \phi^{\tau}
  \right] \Psi[A_{a}^{i}],
  \label{eq:50}
\end{align}
that satisfies
\begin{align}
  \hat{p}_{\tau} \Psi[\phi^{\tau}, A_{a}^{i}]
  = \lambda \Psi[\phi^{\tau}, A_{a}^{i}],
  \label{eq:51}
\end{align}
where $\lambda$ is an unspecified constant.
Note that $\hat{p}_{\tau}$ weakly commutes with every quantum
first-class constraint; therefore, $\hat{p}_{\tau}$ is the physical
observable in the sense of Dirac~\cite{Matschull1996}.

A possible solution of the constraints \eqref{eq:47} and \eqref{eq:48}
is expressed as
\begin{align}
  &\Psi_{\lambda, R}[\phi^{\tau}, A_{a}^{i}]
    = \Psi_{\lambda}[\phi^{\tau}]\Psi_{\lambda, R}[A_{a}^{i}]
    \notag \\
  &= \exp\left[
    \frac{\lambda}{-2\i\hbar k} \int \d^{3}x\ \phi^{\tau}
    \right]
    \exp\left[
    \frac{6}{\i\hbar k \beta^{3}\lambda}
    \left(
    Y_{\text{CS}}[A_{a}^{i}] - 2 \left( 1+\beta^{2} \right)
    \int \Tr ( A \wedge R)
    \right)
    \right],
    \label{eq:52}
\end{align}
where
\begin{align}
  Y_{\text{CS}}[A_{a}^{i}]
  &= \int \Tr \left( A\wedge \d A
    + \frac{2}{3}A\wedge A\wedge A \right)
    \label{eq:53}
\end{align}
is the Chern-Simons functional, $\Tr$ indicates the trace of
SU(2) generators $T_{i}$, which is normalized as
$\Tr(T_{i}T_{j})=-\frac{1}{2}\delta_{ij}$, and
$\lambda$ and $R_{ab}^{i}$ are parameters associated with the
cosmological constant and the spatial curvature, respectively.
We would like to emphasize that the form of the second factor in
\eqref{eq:52}, $\Psi_{\lambda, R}[A_{a}^{i}]$, was originally proposed
in Ref.~\cite{Randono2006} as a generalization of the Kodama state for
real values of $\beta$.

Since this state is a pure phase, we can define a naive inner product
\begin{align}
  \left\langle \Psi_{\lambda', R'} \middle|
  \Psi_{\lambda, R}\right\rangle
  = \int \mathcal{D}\phi^{\tau} \mathcal{D}A\
  \Psi_{\lambda', R'}^{\ast}[\phi^{\tau},A_{a}^{i}]
  \Psi_{\lambda, R}[\phi^{\tau},A_{a}^{i}]
  \sim \delta(\lambda-\lambda')\delta(R-R'),
  \label{eq:54}
\end{align}
where
\begin{align}
  \delta(R-R')=\prod_{x}\prod_{a,b,i}
  \delta\left( R_{ab}^{i}(x)-R'{}_{ab}^{i}(x) \right).
  \label{eq:55}
\end{align}
The inner product \eqref{eq:54} has an undesirable property.
Due to the factor $\delta(R-R')$, when $R_{ab}^{i}$ and
$R'{}_{ab}^{i}$ have different values, this inner product vanishes
even if $R_{ab}^{i}$ and $R'{}_{ab}^{i}$ are connected by the gauge
and spatial diffeomorphism transformations.
We can improve this inner product by using the group averaging
technique \cite{Randono2006}
\begin{align}
  \left( \Psi_{\lambda', R'} \middle| \Psi_{\lambda, R} \right)
  =\int \mathcal{D} g\ \left\langle \Psi_{\lambda', \varphi_{g}R'}
  \middle| \Psi_{\lambda, R} \right\rangle
  \sim \delta(\lambda-\lambda')\int \mathcal{D}g\
  \delta(R- \varphi_{g} R'),
  \label{eq:56}
\end{align}
where $\varphi_{g}$ is the gauge and spatial diffeomorphism
transformations parametrized by $g$, and $\int \mathcal{D}g$ is an
integral over both transformations.
The inner product \eqref{eq:56} does not vanish when
$\lambda=\lambda'$ and $R'{}_{ab}^{i}$ can reach $R_{ab}^{i}$ by
these transformations.
Note that these transformations do not affect $\lambda$.
We find that the state
\begin{align}
  \left( \Psi_{\lambda, R} \right|
  &=\int \mathcal{D}g\ \left\langle
    \Psi_{\lambda, \varphi_{g}R} \right|
    \label{eq:57}
\end{align}
is invariant under the gauge and spatial diffeomorphism
transformations, because
\begin{align}
  \left( \Psi_{\lambda, R} \right| \hat{U}(\varphi_{g'})
  = \int \mathcal{D}g\
  \left\langle \Psi_{\lambda, \varphi_{g'}\circ\varphi_{g}R}\right|
  = \left( \Psi_{\lambda, R} \right|,
  \label{eq:58}
\end{align}
where $\hat{U}(\varphi_{g'})$ is the operator corresponding to these
transformations.
This is an analog of the strategy to obtain the gauge and spatial
diffeomorphism invariant state in loop quantum gravity
\cite{Ashtekar1995,Perez2004}.

One can write the curvature operator $\hat{R}_{ab}^{i}$ by using the
inner product \eqref{eq:56} as
\begin{align}
  \int \d^{3}x\ \tilde{X}_{i}^{ab}\hat{R}_{ab}^{i}
  = \int \d^{3}x\ \tilde{X}_{i}^{ab}
  \int \mathcal{D}g\mathcal{D}R'\mathcal{D}\lambda'\
  \varphi_{g}R'{}_{ab}^{i}
  \left| \Psi_{\lambda', \varphi_{g}R'}\right\rangle
  \left\langle \Psi_{\lambda', \varphi_{g}R'} \right|,
  \label{eq:59}
\end{align}
where $\tilde{X}_{i}^{ab}$ is a densitized test function, and
$\int \mathcal{D}R'$ is an integral over the curvature parameter $R'$
modulo the gauge and spatial diffeomorphism transformations.
From Eq.~\eqref{eq:59}, we have
\begin{align}
  \int \d^{3}x\ \tilde{X}_{i}^{ab}\hat{R}_{ab}^{i}
  \left| \Psi_{\lambda, R} \right\rangle
  = \int \d^{3}x\ \tilde{X}_{i}^{ab}R_{ab}^{i}
  \left| \Psi_{\lambda, R} \right\rangle.
  \label{eq:60}
\end{align}
Using \eqref{eq:51}, \eqref{eq:60}, and
\begin{align}
  \hat{E}_{i}^{a}\Psi_{\lambda,R}[\phi^{\tau},A_{a}^{i}]
  = \frac{3}{\beta^{2}\lambda}\epsilon^{abc}
  \left[ F_{bci}-\left( 1+\beta^{2} \right) R_{bci} \right]
  \Psi_{\lambda,R}[\phi^{\tau},A_{a}^{i}],
  \label{eq:61}
\end{align}
we find that the state $\Psi_{\lambda,R}[\phi^{\tau}, A_{a}^{i}]$
satisfies the Hamiltonian constraint:
\begin{align}
  &\hat{\Phi}'[X] \Psi_{\lambda,R}[\phi^{\tau}, A_{a}^{i}]
    \notag \\
  &= \frac{1}{2k\beta^{2}}\int \d^{3}x\ \frac{X}{\sqrt{\det \hat{E}}}
    \epsilon^{ijk}\hat{E}_{i}^{a}\hat{E}_{j}^{b}
    \left[
    \left( 1+\beta^{2} \right) \hat{R}_{abk}
    -\hat{F}_{abk}
    + \frac{\beta^{2}}{6}\epsilon_{abc}\hat{p}_{\tau}\hat{E}_{k}^{c}
    \right]\Psi_{\lambda,R}[\phi^{\tau}, A_{a}^{i}]
    =0.
    \label{eq:62}
\end{align}
Thus, under the appropriate inner product \eqref{eq:56}, the state
$\left| \Psi_{\lambda, R} \right)$ is a solution of quantum covariant
unimodular gravity.

\section{Conclusions}
\label{sec:conclusions}

In this work, we have analyzed the full theory of covariant unimodular
gravity in terms of the connection representation with real values of
the Barbero-Immirzi parameter $\beta$.
Unlike the original unimodular gravity, the Hamiltonian constraint of
covariant unimodular gravity \eqref{eq:32} is first class.
Therefore, the constraint structure of covariant unimodular
gravity is closer to that of GR than that of the original unimodular
gravity.
The subtle difference from GR is that the cosmological constant in GR
is replaced with the canonical momentum $p_{\tau}$ which is regarded
as a constant of motion and a Dirac observable.

In the original unimodular gravity, the Kodama state is
not a solution of the quantum constraints \cite{Yamashita2020}.
On the other hand, in covariant unimodular gravity, the solution
of the constraints can be obtained by extending the Kodama state.
The state \eqref{eq:52} is regarded as a natural extension of the
state proposed in Ref.~\cite{Randono2006}.
Since this state is a pure phase, it is delta-function normalizable.
In addition, since all the canonical variables are real, we can avoid
the reality condition problem.
The main difference from the quantum states of GR in
Ref.~\cite{Randono2006} is that each state is labeled not only by the
spatial curvature $R_{ab}^{i}$ (modulo the gauge and spatial
diffeomorphism transformations) but also by the cosmological constant.
This implies that a general solution of the physical state can be
written as a superposition of the eigenstates of $\hat{p}_{\tau}$
and $\hat{R}_{ab}^{i}$.
Thus, at least in this framework, covariant unimodular gravity is
different from GR at the quantum level.
Note that such a superposition of the different values of the
cosmological constant has already appeared in previous studies both
within the Hamiltonian formalism \cite{Unruh1989a,Bombelli1991} and
the path integral formalism \cite{Ng1990,Bufalo2015,Smolin2009}.

It is still unclear whether the original unimodular gravity
and covariant unimodular gravity differ at the quantum level.
In the original unimodular gravity, only the wave functional with zero
cosmological constant has been found \cite{Yamashita2020}.
On the other hand, in covariant unimodular gravity,
$\Psi_{\lambda, R}[\phi^{\tau}, A_{a}^{i}]$ \eqref{eq:52} is a state
with a non-vanishing cosmological constant.
To compare the two theories, one needs to find a solution with a
non-vanishing cosmological constant in the original unimodular
gravity.
Note that even if such a state is found, it is not so obvious whether
the difference in the physical states provides different physical
predictions.

Another approach to confirm the difference between the two unimodular
theories is to compare the observables.
In general, a physical observable must commute with every first-class
constraint \cite{Matschull1996}, while the two unimodular theories
have different first-class constraints.
If we can define appropriate observables,
it will be easier to compare the two theories.
Let us consider $V_{\tau}=\int \d^{3}x\, sN\det e$ as an
example of a candidate for the physical observable.
Here, $s=\text{sign}(\det e)$, and the integral is over the spacelike
surface parametrized by $\tau$.
In covariant unimodular gravity, the Poisson bracket between
$V_{\tau}$ and the Hamiltonian constraint does not weakly vanish:
$\left\{ V_{\tau}, \Phi'[X] \right\}
= \int \d^{3}x\, sXN K_{a}^{i}E_{i}^{a} \not\approx 0$.
This implies that, in quantum theory, not all the constraints commute
with $\hat{V}_{\tau}$, and a physical state is not an eigenstate of
$\hat{V}_{\tau}$.
Hence, $V_{\tau}$ and the four-volume $\int \d\tau\, V_{\tau}$ are
not observables in covariant unimodular gravity.
On the other hand, in the original unimodular gravity, the unimodular
constraint $N\det e - \alpha=0$ is a second-class constraint.
This means that $N$ is not an independent variable, and can be
eliminated as $N=\alpha (\det e)^{-1}$.
Therefore, $V_{\tau}=\int \d^{3}x\, s\alpha$.
In this case, the Poisson bracket between $V_{\tau}$ and every
first-class constraint weakly vanishes.
However, $V_{\tau}$ and the four-volume in this theory should be
regarded as constants associated with the Hamiltonian and the
Lagrangian rather than observables.
Thus, $V_{\tau}$ is not an appropriate quantity to compare the
two theories.
To confirm the (in)equivalence of the two theories,
one needs to find some other quantities that can be physical
observables.

It is worthwhile to investigate how covariant unimodular gravity is
extended into loop quantum gravity.
Both covariant unimodular gravity discussed above and loop quantum
gravity employ real values of the Barbero-Immirzi parameter $\beta$.
Therefore, one can expect that this theory can be naturally extended
into a full theory of loop quantum gravity.
Note that the symmetry-reduced models of covariant unimodular
gravity in the context of loop quantum cosmology has been studied
in Refs.~\cite{Chiou2010,Bunao2017,Sartini2021}.

Another direction for future research is to investigate the relation
between the extended Kodama state \eqref{eq:52} and other physical
states of unimodular gravity.
In Refs.~\cite{Magueijo2020,Alexander2021}, the Hartle-Hawking state,
which is the solution of quantum GR in the ADM representation,
is interpreted as the Fourier dual of the Kodama state.
On the other hand, there is also an argument that it is difficult to
translate the Hartle-Hawking state straightforwardly into the
connection representation \cite{Dhandhukiya2017}.
The question of how these arguments are modified in covariant
unimodular gravity is left for future study.

\section*{Acknowledgments}
The author is grateful to Makoto Fukuda for useful comments on the
manuscript.

{\small
  \providecommand{\href}[2]{#2}
  \begingroup
  \raggedright\endgroup
}
\end{document}